\documentclass[12pt]{article}
\usepackage{graphicx}  
\usepackage{dcolumn}   
\usepackage{bm}        
\usepackage{amssymb}   
\usepackage{epstopdf}
\usepackage{setspace}
\usepackage[margin=1in]{geometry}

\newcommand{\textss}[1]{\scriptsize \mbox{#1}}

\newcommand{\ODmin}{2 \times 10^{-10}}
\newcommand{\ODHz}{1 \times 10^{-9}/\sqrt{\mbox{Hz}}}
\newcommand{\ODSamplemin}{10^{-8}}
\newcommand{\ColumnDensity}{ $10^{10}$ molecules/cm$^2$}
\newcommand{\Finesse}{370}

\begin{document}

\section*{}
\begin{center}
{\large \textbf{Cavity-Enhanced Ultrafast Transient Absorption Spectroscopy} \par}

Melanie A. R. Reber$^\diamond$ \let\thefootnote\relax\footnote{$^\diamond$ these authors contributed equally to this work}, Yuning Chen$^\diamond$ $\&$ Thomas K. Allison$^*$

\textit{Stony Brook University, Stony Brook, NY 11794-3400}

\end{center}

\section*{}
\textbf{We present a new technique using a frequency comb laser and optical cavities for performing ultrafast transient absorption spectroscopy with improved sensitivity. Resonantly enhancing the probe pulses, we demonstrate a sensitivity of $\Delta$OD $ = \ODHz$ for averaging times as long as 30 s per delay point ($\Delta$OD$_{\textss{min}} = \ODmin$). Resonantly enhancing the pump pulses allows us to produce a high excitation fraction at high repetition-rate, so that signals can be recorded from samples with optical densities as low as OD  $\approx \ODSamplemin$, or column densities $<$\ColumnDensity. This high sensitivity enables new directions for ultrafast spectroscopy.}

\section*{}

The advent of the mode-locked Ti:Sapphire laser in the early 1990's \cite{Spence_OptLett1991} made ultrafast pump-probe measurements routine and widely accessible to a broad range of scientists. This was largely due to the Ti:Sapphire laser's robustness compared to the dye-lasers it replaced. However, another aspect of the Ti:Sapphire laser that made many experiments possible was its capability of low-noise performance. This allowed for measurements of very small changes in the optical properties of a sample induced by a pump pulse, which is necessary when the sample is either dilute \cite{Schriever_RSI2008, Boschetto_NanoLett2013, Rottger_AppPhysB2015} or must be excited weakly to probe the desired physics \cite{Basov_RMP2011,Orenstein_PhysToday2012}. Even with relatively noisy chirped pulse amplified systems, one can measure changes in absorbance or reflection to a few parts in $10^6$, and ultrafast transient absorption spectroscopy is the simplest and most widely applied form of ultrafast spectroscopy \cite{Berera_PhotoSynthRes2009}.

Despite this enormous progress, there remain many samples for which ultrafast optical spectroscopy is still prohibitively difficult.  Most directly related to the current work are the ``designer" gas-phase molecules and molecular clusters that can be produced in a supersonic expansion. With optical spectroscopy seemingly hopeless, ultrafast experimenters perform measurements on these systems by ionizing the molecules with UV pulses or strong fields \cite{Stolow_ChemRev2004,Kosma_JPhysChemA2008}, detecting the resulting ions and electrons. This is indeed extremely sensitive due to the capabilities of single particle detection and background free signals. However, ionization projects the molecular state  of interest onto a very different manifold of final states than optical measurements, and this can make the comparison of experimental data from gas phase and condensed phase highly non-trivial \cite{Kohler_AnnRevPhysChem2009,Saigusa_JPhotoChemBio2007}. Furthermore, while dynamics of electronically excited states can be probed by ionization, there exists no ionization-based methods for probing purely vibrational dynamics analogous to the powerful tools of ultrafast infrared spectroscopy \cite{Zanni_Book2011}.

It was realized in the early days of lasers that an optical resonator is useful for absorption enhancement due to the fact that light passes through a sample many times \cite{Hansch_IEEE1972}. Today, cavities are widely used in many different contexts for the enhancement of optical signals \cite{Gagliardi_Book2013}. Recently, several groups have employed the cavity-enhancement of frequency combs \cite{Jones_OptLett2002, Jones_OptLett2004} for performing spectroscopy. Thinking of the comb as ``a million stable lasers at once" \cite{Hall_IEEE2001}, one can record spectra that are simultaneously sensitive, broad-band, and high-resolution \cite{Adler_AnnRevChem2010, Gohle_PRL2007}. These techniques, called ``Cavity-Enhanced Direct Frequency Comb Spectroscopy" (CE-DFCS), have now been applied to rapid trace gas detection \cite{Foltynowicz_APB2013,Bernhardt_NatPhot2010}, breath analysis \cite{Foltynowicz_APB2013,Thorpe_OptExp2008}, and microsecond time-resolved kinetics \cite{Fleisher_JPCL2014}. However, this previous spectroscopy work has largely neglected the pulsed nature of the intracavity comb, and that these pulses can be used for the sensitive detection of ultrafast time-resolved signals \cite{Benko_PRL2015, Jones_CLEO2015}. Indeed, as we show here, the gains for nonlinear spectroscopy can actually be larger, since both pump and probe pulses can be resonantly enhanced.

In this article, we report ultra-sensitive transient absorption measurements with 120 fs resolution using resonant enhancement of pump and probe pulse trains at 87 MHz repetition rate. We demonstrate measurements with a noise level as low as $\Delta$OD = $\ODmin$. To our knowledge, this represents an improvement of more than 5,000 over previous transient absorption results \cite{Schriever_RSI2008}, and a 100-fold improvement over previous high-performance time-resolved reflectivity measurements \cite{Boschetto_NanoLett2013}. Importantly, by also resonantly enhancing the pump pulses to high average power ($\sim$ 50 W), there is no compromise between the $\Delta$OD detection limit and the fraction of molecules that can be excited, so that the large sensitivity improvement translates directly into the usable molecular concentration. This work extends ultrafast optical spectroscopy to molecular beams, and also can be adapted to other applications where higher sensitivity is needed \cite{Basov_RMP2011,Liu_PRB2011}.

\section*{Principle of Ultrafast Signal Enhancement}

It may seem contradictory that a high-finesse optical cavity, with long coherence time, can enhance ultrafast signals which decohere rapidly, but the fundamental mechanism for ultrafast signal enhancement is the same as in CE-DFCS and other cavity-enhanced spectroscopies. In CE-DFCS, an intracavity pulse traverses a sample of molecules, identical each round trip in the limit of weak excitation, many times. The signal is enhanced by  $\mathcal{F}/\pi$ for the case of an impedance matched ring cavity, where $\mathcal{F}$ is the cavity finesse \cite{Gagliardi_Book2013}. In cavity-enhanced transient absorption spectroscopy (CE-TAS), the probe pulse also traverses a sample of molecules many times, but now we prepare this sample in an excited state using a pump pulse. This excitation, done at a repetition rate equal to the cavity's free spectral range, is identical every round trip, so from the point of view of the probe pulse absorption, there is no difference between CE-DFCS and CE-TAS, and the resulting signal enhancement is the same. Time resolution comes from the time dependence of the excited state of the molecules, as in a normal pump-probe experiment, and the time-resolved signal is recorded by simply varying the pump-probe delay with an external translation stage. Viewed in the frequency domain, CE-TAS uses both the intracavity comb's spectral amplitude and spectral phase, which encodes the pulse shape and delay, whereas CE-DFCS uses only the amplitude.

\begin{figure}
    \centering
    \includegraphics[width=6.5in]{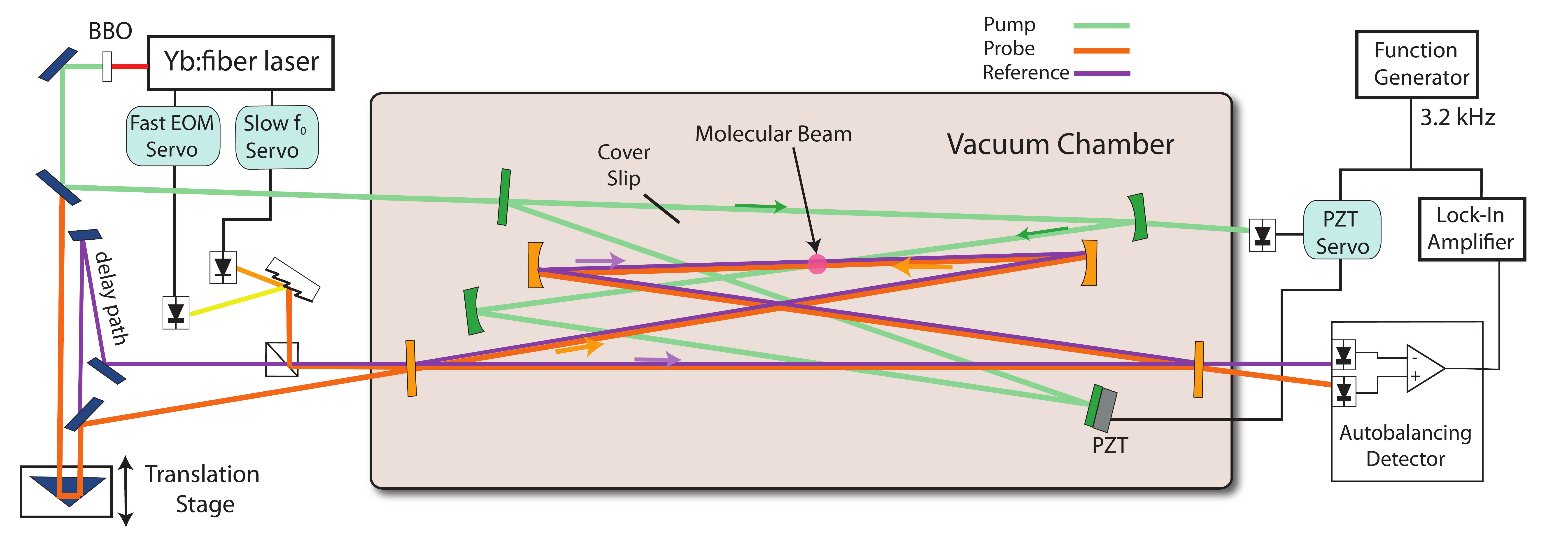}
    \caption{\textbf{Schematic of the CE-TAS system}. Ultrafast transient absorption experiments are performed in a molecular beam at the common focus of two optical resonators, one for the pump pulses and another for the probe pulses. Delayed counter-propagating reference pulses are used for common mode noise subtraction. The beams are color coded for clarity, but are all the same wavelength in the current experiment. More details are described in the main text and the methods section.}
    \label{fig:Apparatus}
\end{figure}

A diagram of our CE-TAS implementation is shown in figure \ref{fig:Apparatus}. The foci of two femtosecond enhancement cavities (fsEC's) cross at an angle of $\sim 20$ mrad above a nozzle where sample molecules are introduced in a supersonic expansion. Pump and probe pulses traverse the sample in the same direction to avoid broadening of the temporal resolution due to the transit time through the sample. Resonant enhancement of the pump pulses, with a lower finesse cavity, provides passive amplification such that a large ($\sim 1\%$) fraction of the sample molecules can be excited at high repetition rate. High frequency modulation/demodulation techniques can still be employed without penalty as long as the modulation frequency is substantially lower than the cavity linewidths, but two complications arise in detecting the signal on the intracavity probe light. The first is that while the probe cavity enhances the signal, it also increases the amplitude noise on the transmitted probe light since the probe cavity turns the laser's frequency noise into amplitude noise. This is commonly encountered in cavity-enhanced spectroscopy \cite{Gagliardi_Book2013}. The second is that the supersonic expansion flow speed is not fast enough to replenish the sample within one cavity round trip, so that the sample is reused for approximately 3-10 pump-probe sequences, and the probe pulse measures not only femtosecond signals from the excited population immediately preceding it, but also nanosecond signals from several preceding pump pulses. This problem is unique to CE-TAS. Coherent molecular motion can be suppressed on nanosecond time scales by collisions with carrier gas \cite{Zewail_JPhotoChemPhotoBio1992} or the natural decay of the excited state (the likely case for most molecules of interest to ultrafast spectroscopy), but still a large ground-state bleach signal is expected to persist.

\begin{figure}
    \centering
    \includegraphics[width=3.25in]{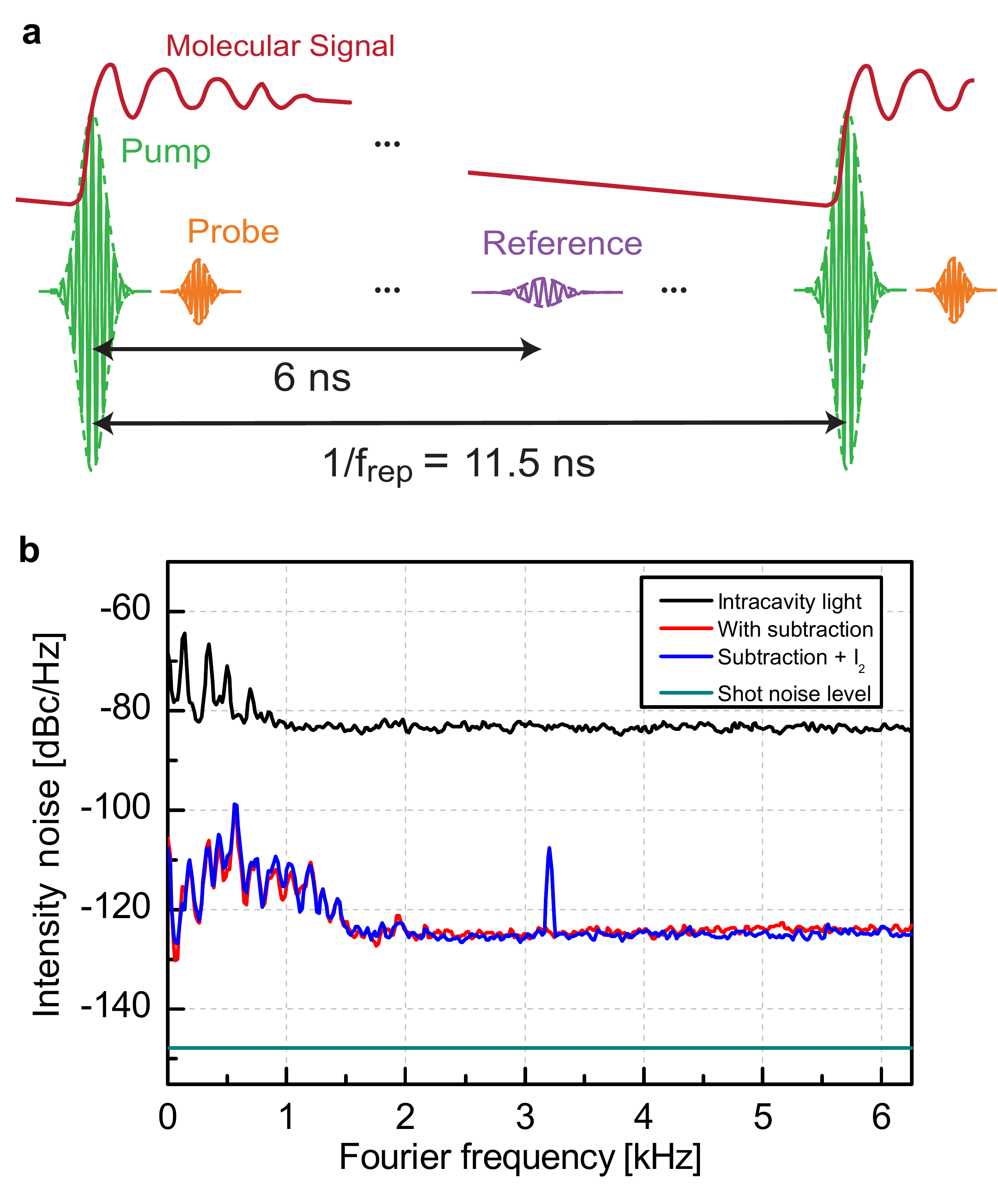}
    \caption{ \textbf{Noise subtraction}. \textbf{a}, Pulse sequence at the molecular sample. The probe and reference share common mode noise but sample different molecular signals. \textbf{b}, Intracavity relative intensity noise (RIN) spectrum with and without subtraction of the reference pulse train. More than 40 dB of RIN can be suppressed. With the introduction of sample molecules, a coherent spike at the pump modulation frequency of $3.2$ kHz is observed.}
    \label{fig:noise}
\end{figure}

We address both of these complications by coupling a counter-propagating reference pulse train to the probe cavity and recording the difference between probe and reference pulses in an autobalanced detection scheme.  The pulse sequence at the molecular sample is illustrated in figure \ref{fig:noise}a. The pump and reference pulses share common mode noise, but different signals. The probe pulse arrives shortly after the pump and records the femtosecond signal of interest, while the reference pulse arrives 6 ns later and samples the persistent bleach signal.  Counter-propagation allows the reference beam to be easily separated from the probe beam in the ring cavity geometry and also reduces the concern for any parasitic coherent excitation it might produce, since this is effectively smeared out in time. Noise reduction is shown in figure \ref{fig:noise}b, where subtraction of the reference pulse reduces the relative intensity noise (RIN) on the intracavity light by more than 40 dB at the modulation frequency, allowing small signals to be recovered.

\section*{Two-Cavity Operation}

In our CE-TAS system (figure \ref{fig:Apparatus}), the probe cavity consists of a nearly impedance matched four-mirror bow-tie ring cavity with two 50 cm radius of curvature concave mirrors for a finesse of $\mathcal{F}_{\textss{probe}} =  \Finesse$  and an absorption enhancement factor of approximately 120. We measured the probe cavity finesse using a ring-down technique, as described in \cite{Thorpe_OptExp2008}. The intracavity focus is calculated to be 70 $\mu$m FWHM. We use a two point Pound-Drever-Hall locking scheme to lock the comb to the cavity, as described in \cite{Foltynowicz_PRL2011}. A tight lock of the center of the frequency comb to the probe cavity is achieved using an electro-optic modulator (EOM) inside the Yb:fiber oscillator cavity. A slower feedback loop moves a diffraction grating inside the oscillator to adjust the laser's carrier-envelope offset frequency ($f_0$) to match that of the probe cavity, which is determined by the dispersion of the probe cavity optics \cite{Jones_OptLett2002}.

The pump cavity is a four-mirror bow-tie ring cavity with a 110 $\mu$m focus formed by 75 cm radius of curvature mirrors. The pump cavity is strongly overcoupled \cite{Nagourney_Book2010}, with the loss dominated by the 3\% transmission of the input coupler. Since we have two cavities but the laser has only one $f_0$, determined by the lock to the probe cavity, we adjust the $f_0$ of the pump cavity to match that of the probe. This is achieved by inserting a $\sim 150$ $\mu$m fused silica microscope cover slip at Brewster's angle inside the pump cavity. Different cover slips were tried until one with the correct thickness was found. The pump cavity is locked to the comb using a fast Piezo-electric transducer (PZT) \cite{Briles_OptExp2010} and ``side-of-line" lock where the cavity's transmitted power is used as the error signal. The pump power is then modulated at 3.2 kHz by adjusting the lock-point and providing a feed-forward signal to the PZT in concert. Calculations indicate that for the low finesse ($\sim 200$) employed here, when the pump cavity's $f_0$ is the same as the laser's, changes in the intracavity pulse shape and delay are negligible as the intracavity pump power is modulated in this way. We have experimentally verified this by recording signals with different modulation depths and DC offsets to the lock-point, confirming that the observed signal scales simply as the product of the intracavity power and the modulation depth without changing shape.

\section*{Transient Absorption Measurements}

For this demonstration, we chose to study gas-phase molecular iodine (I$_2$) excited to the $B$ $^3\Pi_{0_{u}}^+$ state. The $B$ state of I$_2$ has been extensively studied using both time-resolved \cite{Zewail_JPhotoChemPhotoBio1992} and static \cite{Mulliken_JChemPhys1971,Steinfeld_JCP1965} spectroscopy, and thus is a good candidate for testing this new technique. I$_2$ is introduced at the common focus in a continuous He-seeded supersonic expansion. Pump and probe frequency combs centered at 529 nm are produced via second harmonic generation of a home-built 87 MHz Yb:fiber laser system with a $\sim 30$ kHz comb tooth linewidth. More details on this laser, the molecular beam source, and the vacuum system can be found in the methods section.  In our experiment with pump and probe at the same wavelength, we expect to observe a bleach of the ground state absorption along with stimulated emission occurring when the excited state wavepacket returns to the Franck-Condon region.

\begin{figure}
    \centering
    \includegraphics[width=6.5in]{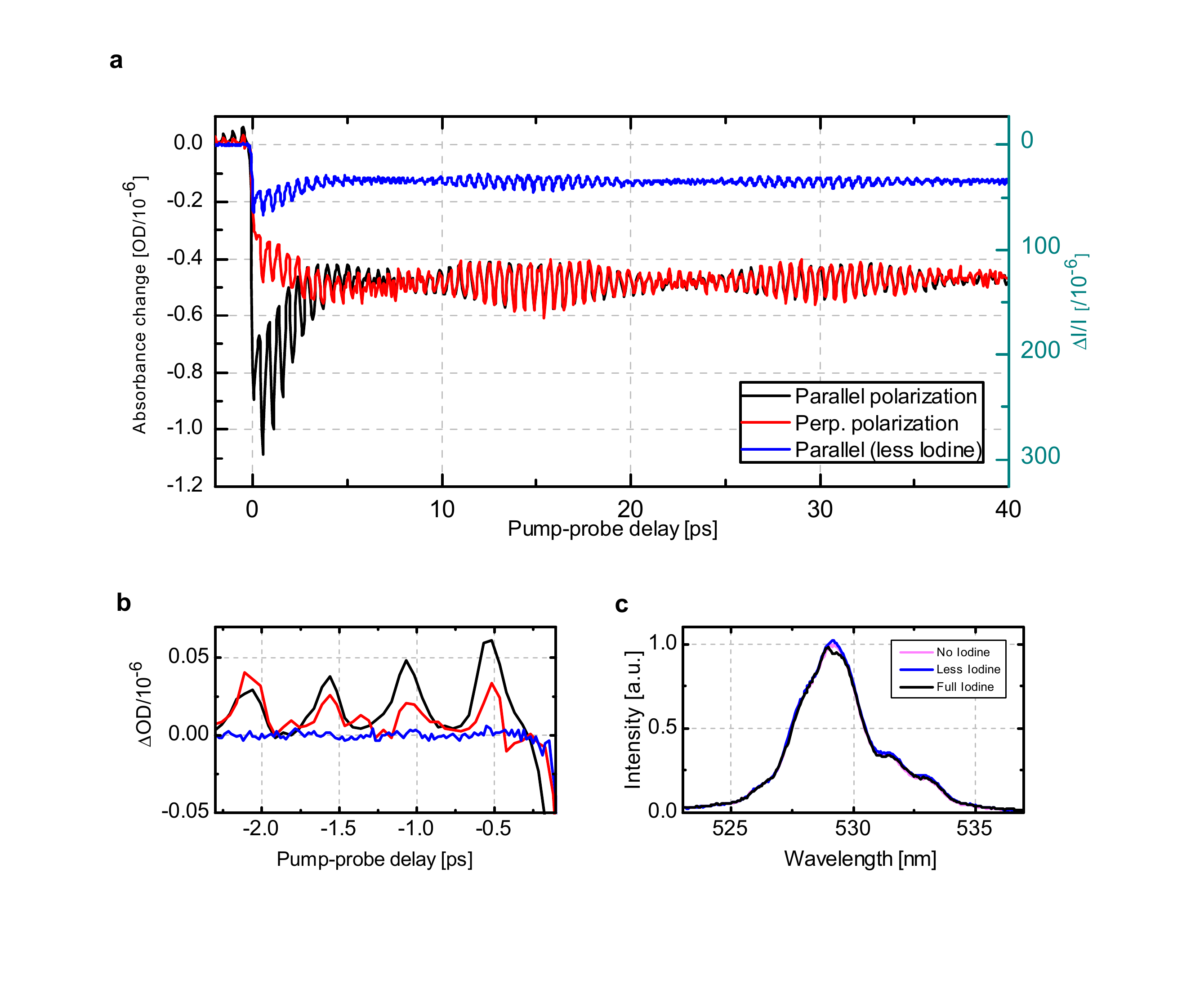}
    \caption{ \textbf{Transient absorption data.} \textbf{a}, Measurements of a molecular wavepacket in the $B$ $^3\Pi_{0_{u}}^+$ state of I$_2$. Stimulated emission occurs when the molecule returns to the Franck-Condon region. Three vibrational states on the $B$ state surface are predominantly excited, giving rise to the observed vibrational beating pattern. Rotational motion causes a rapid decay of the polarization anisotropy. The perpendicular polarization data was taken under different conditions for the pump cavity and has been multiplied by $3.2$. \textbf{b}, If the partial pressure of I$_2$ in the vacuum chamber gets too high, an artifact of the CE-TAS scheme is visible before time zero due to distortion of the intracavity pulses. This artifact is effectively eliminated by reducing the gas flow. \textbf{c}, The intracavity light spectrum is also visibly distorted when the artifact appears.}
    \label{fig:longsignal}
\end{figure}

Data is recorded by scanning the external delay stage and recording the subtracted signal with a lock-in amplifier. Typical pump-probe traces for both parallel and perpendicular polarizations are shown in figure \ref{fig:longsignal}a. Coherent oscillations of the $B$-state wavepacket are observed on top of a ground state bleach signal. The observed oscillation frequency, vibrational beat pattern, and rotational dephasing \cite{Felker_JPhysChem1986, Weiner_book2009} are in accordance with what is expected based on the known spectroscopy of iodine \cite{Steinfeld_JCP1965,Zewail_JPhotoChemPhotoBio1992} and our laser spectrum. The right y-axis shows the fractional change in the transmitted light intensity ($\Delta I/ I$) determined from the DC photocurrent at the detector and the calibrated lock-in gain. The left y-axis shows the change in the molecular ensemble's optical density ($\Delta OD$), calculated from $\Delta I/ I$ and the measured cavity finesse via
\begin{equation}
\Delta \mbox{OD} = -\log_{10}(e) \frac{\pi}{ \mathcal{F}_{\textss{probe}} } \left( \frac{\Delta I}{I} \right)
\end{equation}
Systematic uncertainty regarding these quantities is estimated to contribute less than 20\% error to the y-axis calibration. Diffraction effects due to spatial inhomogeneity in the excited sample (e.g. from pump misalignment) would only increase the signal enhancement factor \cite{Pupeza_NatPhot2013}. For long delay scans, CE-TAS enjoys a practical convenience that pump/probe overlap is determined solely by the spatial modes of the pump and probe fsEC's, and is therefore insensitive to the translation stage position or alignment. If high-pressure argon is used as the carrier gas, a large coherent transient with a FWHM of 120 fs is observed before the oscillatory signal, and this is used to determine the zero of the delay axis and estimate the temporal resolution. We have verified that the raw signal ($\Delta I$) is linear in the pump and probe/reference powers by varying these over one order of magnitude and a factor of 2, respectively.

An additional concern with the CE-TAS scheme arises from the potential distortion of the intra-cavity pulse shape due to absorption from background gas in the vacuum chamber. This is particularly true for a molecule like I$_2$, with sharp spectral features that can cause the probe pulse to develop a wake. Indeed, when flowing large amount of I$_2$ gas, we observe a small oscillatory signal before time zero due to this wake. The appearance of this signal coincides with the onset of I$_2$ absorption lines becoming barely visible in the intracavity light optical spectrum (figure \ref{fig:longsignal}c). This artifact becomes negligible when the I$_2$ flow is reduced, as shown in figure \ref{fig:longsignal}b. As in other forms of cavity-enhanced spectroscopy, the total absorption of the analyte should be kept small with respect to the cavity mirror losses to avoid distortions of the signal.

\begin{figure}
    \centering
    \includegraphics[width=3.25in]{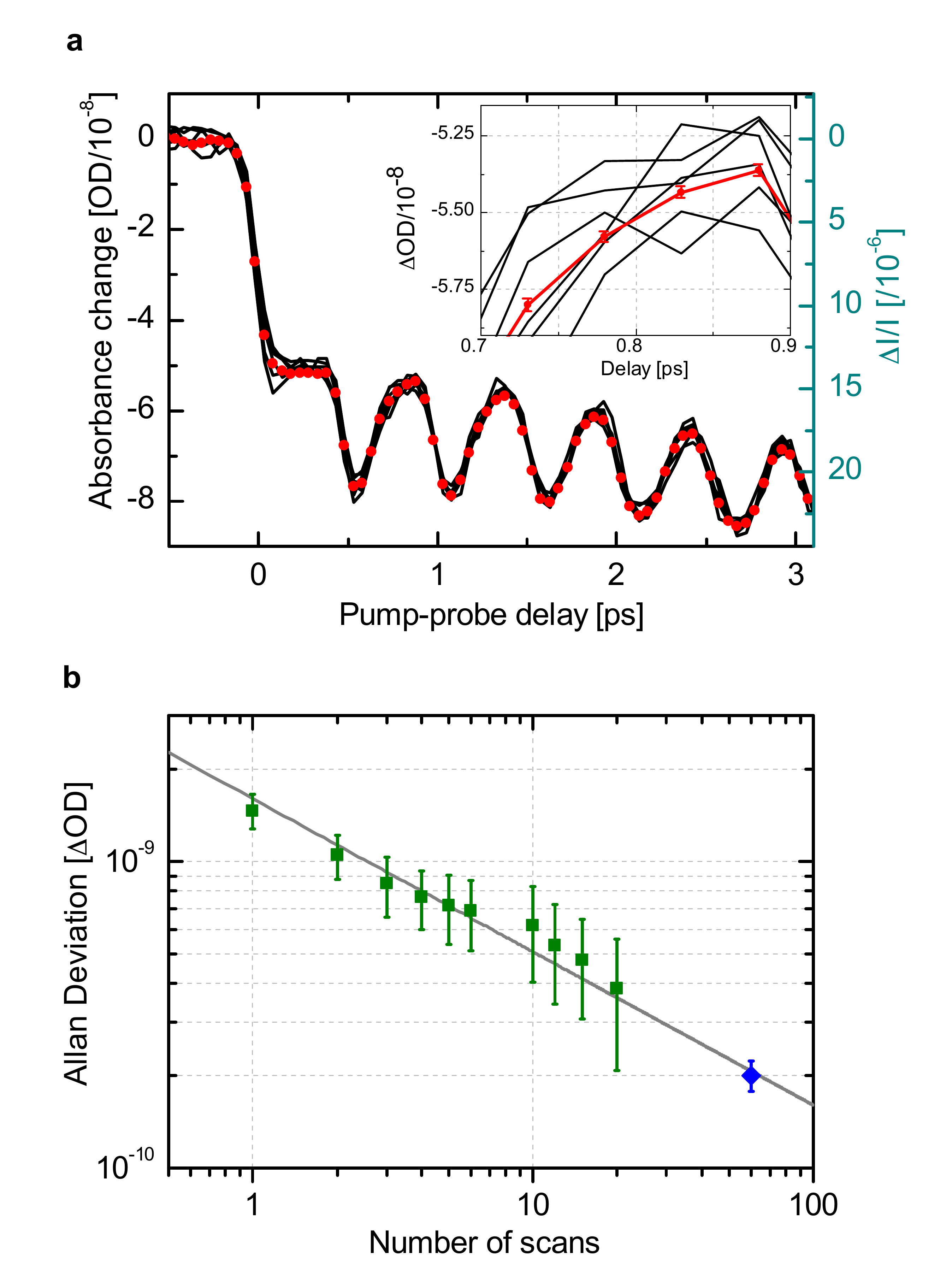}
    \caption{\textbf{Noise performance of CE-TAS}. \textbf{a}, Transient absorption measurements taken with reduced gas flow and perpendicular polarizations. The red dots represent the average of 60 consecutive scans taken over a 1 hour period. Black curves are every 10th scan from the data set. Inset: Zoom-in around 0.8 ps delay. Error bars represent the uncertainty in the mean. \textbf{b}, The green squares show the average of the Allan deviations obtained independently for each delay point. Error bars here are the standard deviation (not the uncertainty in the mean) of this ensemble, to represent the spread in the data. The blue diamond is the average of the error bars of \textbf{a}, along with their standard deviation. The grey line has a slope of -1/2 on the log-log plot, the expected slope for white noise performance.}
    \label{fig:allan}
\end{figure}

We examined the noise performance of CE-TAS by reducing the I$_2$ flow as low as we stably could with our current gas-handling system and recording 60 consecutive scans over a 1 hour period. Each scan, the data are taken for 0.5 s per delay point with perpendicular polarizations and 50 fs step size, for a total accumulation of 30 s per point. Figure \ref{fig:allan} shows every 10th scan along with the average of the complete data set. The error bars in figure \ref{fig:allan}a are the standard deviation of the mean, calculated as the standard deviation of the 60 consecutive measurements, divided by $\sqrt{60}$. The error bars have a mean size of $\Delta \mbox{OD} = 2.0 \times 10^{-10}$ averaged over all delays with a standard deviation of $2 \times 10^{-11}$. These measurements are consistent with the RIN data of figure \ref{fig:noise} and the measurement bandwidth, when the 3 dB difference of phase-sensitive lock-in detection is accounted for. This suggests that even over 1 hour data acquisition times, the measurement is dominated by white noise processes, which is also supported by an Allan deviation analysis, shown in figure \ref{fig:allan}b. We thus report the sensitivity as $\Delta \mbox{OD} = \ODHz$, and our demonstrated detection limit in a practical transient absorption experiment, where averaging time must be distributed over many delay points, as $\Delta \mbox{OD} = \ODmin$.

\section*{Discussion}

The ability to average for long times, due to high frequency modulation/demodulation and the noise canceling scheme unique to CE-TAS, is remarkable when compared to the performance of other cavity-enhnaced spectroscopies, which often reach a flicker noise floor within a few minutes. Still, the current measurement remains technical noise limited. Based on previous work in suppressing noise in CE-DFCS \cite{Foltynowicz_PRL2011,Khodabakhsh_OptLett2014}, achieving shot-noise limited detection should be possible, which would reduce detection limit of the current system by one order of magnitude. Furthermore, the current probe cavity finesse of $\Finesse$ is quite modest for cavity-enhanced spectroscopy, and this can also be improved along with the time-resolution and probe pulse bandwidths that can be achieved \cite{Holzberger_OptLett2015,Gohle_Nature2005}. The methods can be extended to the UV and infrared \cite{Schliesser_NatPhot2012}, the probe light can be spectrally resolved as in conventional transient absorption spectroscopy, and multidimensional spectroscopy can be performed via phase cycling methods \cite{Shim_PCCP2009}.

Even with the current performance, CE-TAS easily extends all-optical ultrafast spectroscopy to a vast array of interesting systems that can only be produced in supersonic expansions. Assuming that pump-induced changes in the absorption are on the same order of magnitude of the ground state absorption, exciting 2\% of the molecules, one can work with samples with optical density as small as 10$^{-8}$. For I$_2$, with absorption cross section of $\sigma \sim 3 \times 10^{-18}$ cm$^2$ \cite{Saiz-Lopez_AtmChemPhys2004}, this translates to a column density less than $10^{10}$ molecules/cm$^2$. Small gas-phase water clusters, for example, can easily be produced with column densities and optical densities much larger than this \cite{Paul_JPCA1997}. Reducing the detection limit further, as discussed above, and increasing the interaction length, could potentially allow for measurements on trapped mass-selected ion clusters \cite{Li_JAmSocMassSpec1998,Majima_PRA2012}.

CE-TAS with only a probe cavity could also benefit other optical measurements. For example, correlated electron systems in condensed matter at low temperature \cite{Basov_RMP2011,Orenstein_PhysToday2012} must be excited very weakly to avoid undesired thermal effects \cite{Liu_PRB2011}. It is in principle possible to incorporate a solid sample into the probe cavity either as a component of a mirror coating or as a wafer at Brewster's angle.

\section*{Methods}

\subsection*{Yb:fiber laser}

Experiments are performed with a home-built Yb:fiber laser system consisting of a passively mode-locked fiber oscillator, a pulse stretcher, a fiber amplifier and a pulse compressor. The oscillator is similar to the one presented in ref. \cite{Nugent-Glandorf_OptLett2011} and a representative schematic can be found there. An intracavity electro-optic modulator (EOM) allows for high speed adjustment of the effective cavity length and tight locking to the probe fsEC. The oscillator's net cavity group delay dispersion (GDD) is tuned by adjusting the distance between the oscillator gratings. We found near zero GDD, but slightly anomalous, gave the quietist operation. We characterized the frequency noise of the oscillator by observing heterodyne beats between the comb teeth and a 1 kHz linewidth CW Nd:YAG laser (Innolight Mephisto), finding the free-running comb tooth linewidth to be approximately 30 kHz. Linear chirped-pulse amplification up to 9 W average power is done in similar fashion to Schibli \emph{et al.} \cite{Schibli_NatPhot2008}, using an anomalous third order dispersion fiber stretcher, an end pumped 5 m long Yb-doped photonic crystal fiber amplifier, and a pair of 1250 line/mm transmission gratings.  

\subsection*{Molecular beam and vacuum system}

Both pump and probe cavities are mounted on an optical platform housed in a 2 ft. by 4 ft. rectangular vacuum chamber. For introducing the I$_2$ sample, He or Ar carrier gas is passed through a room-temperature teflon pick-up cell containing glass wool coated in iodine powder and then expanded into vacuum through a 700 $\mu$m diameter nozzle. To decrease the amount of I$_2$ in the experiment, we merge this flow with a separate stream of pure carrier gas that bypasses the pickup cell, diluting the I$_2$ concentration. All components in the gas handling system downstream of the pickup cell, including the nozzle, are made of teflon to prevent undesired chemistry. The stagnation pressure was varied between 200 Torr and 760 Torr.

The laser beams cross approximately 1 mm above the nozzle. To reduce the background I$_2$ pressure in the chamber, the molecular beam is aimed into a funnel with a 12 mm aperture approximately 5 mm from the nozzle, channeling the gas flow towards a 230 L/s Roots pumping system. Most of the gas load is handled by this Roots pump, and the chamber pressure is maintained below 7 mTorr by a 900 L/s turbo-molecular pump. The partial pressure of I$_2$ in the chamber is further reduced by a dry-ice cooled cold trap, which cryo-pumps the I$_2$ vapor, but not the carrier gas. 

\subsection*{Alignment and polarization control}

Overlap of pump and probe beams at the sample is obtained by aligning both cavity modes through a 100 $\mu$m diameter pinhole placed near the plane of the nozzle. This alignment cannot be optimized in-situ with the current mechanical design, but this could be accomplished by motorizing the pump cavity mirrors. The pump beam polarization is rotated by inserting a wave plate in the pump beam and re-orienting the intracavity cover slip.

\section*{Acknowledgements}
We thank C. J. Johnson, M. K. Liu, and M. Y. Sfeir for helpful discussions regarding ion traps, quantum materials, and ultrafast spectroscopy. We thank T. J. Sears and M. G. White for critical equipment loans. This work was supported by the National Science Foundation under award number 1404296.

\section*{Author contributions}
 All three authors conceived of and designed the experiment, recorded data, analyzed data, and edited the manuscript.

 \section*{Competing interests}
 The authors declare that they have no competing financial interests.

 \section*{Correspondence}
 Correspondence and requests for materials should be addressed to T.K.A. (email: thomas.allison@stonybrook.edu).


\newpage

\newpage

\newpage

\newpage

\end{document}